\documentclass[	twocolumn,
				floatfix,
				showpacs,
				superscriptaddress,
				prb]{revtex4}

\usepackage{graphicx}
\usepackage{amsmath}

\begin{document}

\title{Dirac-Kronig-Penney model for strain-engineered graphene}
	\author{S. Gattenl\"ohner}
	\affiliation{School of Engineering \& Physical Sciences, Heriot-Watt University, Edinburgh EH14 4AS, UK}
	
	\author{W. Belzig}
	\affiliation{Department of Physics, Konstanz University, D--78457 Konstanz, Germany}

	\author{M. Titov}
	\affiliation{School of Engineering \& Physical Sciences, Heriot-Watt University, Edinburgh EH14 4AS, UK}
	\affiliation{Institut f\"ur Nanotechnologie, Karlsruhe Institute of Technology, 76021 Karlsruhe, Germany}

\date{\today}
\begin{abstract}
  Motivated by recent proposals on strain-engineering of graphene
  electronic circuits we calculate conductivity, shot-noise and the
  density of states in periodically deformed graphene.  We provide
  the solution to the Dirac-Kronig-Penney model, which describes
  the phase-coherent transport in clean monolayer samples with
  an one-dimensional modulation of the strain and the electrostatic
  potentials. We compare the exact results to a qualitative
  band-structure analysis. We find that periodic strains induce large pseudo-gaps and suppress charge
  transport in the direction of strain modulation. The strain-induced
  minima in the gate-voltage dependence of the conductivity characterize
  the quality of graphene superstructures.  The effect is especially
  strong if the variation of inter-atomic distance exceeds the value
  $a^2/\ell$, where $a$ is the lattice spacing of free graphene and
  $\ell$ is the period of the superlattice.  A similar effect induced by
  a periodic electrostatic potential is weakened due to Klein
  tunnelling.
\end{abstract}
\maketitle

\section{Introduction}\label{introduction}

Graphene is recognized as the only two-dimensional crystal 
that withstands large deformations and high temperatures 
and is readily integrated into electronic circuits.\cite{Geim09,Neto09,Geim07} 
The high electron mobility and unique spectral characteristics single out graphene
as a promising key element for electronic and optoelectronic 
applications.\cite{Meyer07,Bolotin08,Ponomarenko08,Lin10,Kim09,Wang08} 

The well-known Dirac dispersion of electronic excitations in graphene 
is responsible for a number of interesting analogies with relativistic 
quantum electrodynamics,\cite{Katsnelson07} but prevents the use of graphene 
in field-effect transistors due to the absence of energy band-gaps.\cite{Novoselov04,Novoselov05} 
The strain-engineering of graphene circuits has been recently 
suggested as a means to bypass this difficulty.\cite{Pereira2009a,Guinea10a} 
In the Dirac-fermion picture of graphene, the strain gradient is equivalent 
to the presence of a pseudo-magnetic field, which can be manipulated 
to induce a zero-field quantum Hall effect and a topological isolator state.\cite{Guinea10b}
The variation of strain may also induce substantial variation in the on-site electron energies,
however, this effect is reduced by screening.\cite{Oppen09}

The experimental realizations of strained graphene confirm its surprisingly 
high elasticity and pave the way for the development of strain-engineered 
graphene electronics.\cite{Bunch08,Kim09} Different techniques have been proposed 
to produce graphene samples with controlled periodical variations of strain.\cite{Bao09,Pletikosic09,Parga08} 
First transport measurements of graphene superstructures
have been already reported.\cite{Teague09} Below we analyze 
the phase-coherent charge transport in periodically strained graphene samples 
and propose a way to characterize the quality of graphene superstructures on the basis 
of their transport properties. 

More specifically we use the exact solution of the Dirac-Kronig-Penney model to calculate 
density of states, conductance, and shot noise in transport through finite size graphene 
samples with periodic potentials. The scattering off the metal leads is taken into account 
in all quantities. The position, the width, and the shape of the conductance minimum 
as well as the shot noise maximum associated with the periodic superstructure are analyzed in detail. 

\begin{figure}[tb]
\centerline{\includegraphics[width=\columnwidth]{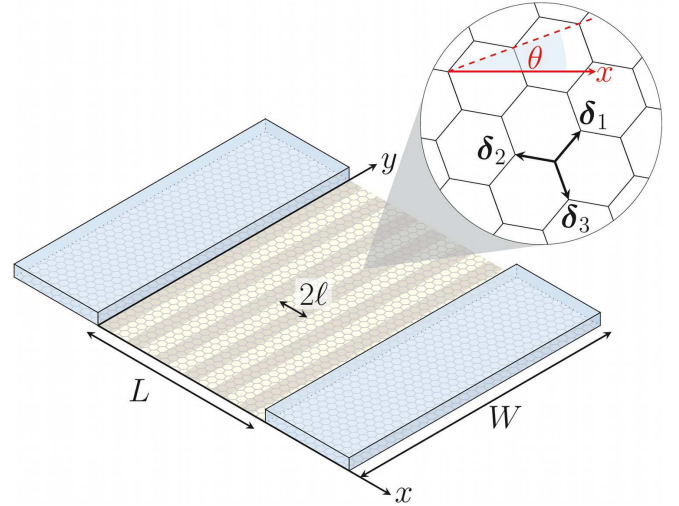}}
\caption{(Color online) Schematic illustration of the graphene setup with metal leads for $x<0$ and $x>L$. 
The angle $\theta$ specifies the orientation of the honeycomb lattice with respect to the transport direction $x$.}
\label{fig:theta}
\end{figure}

This paper is organized as follows. In Section \ref{sec:model} 
the effective Dirac Hamiltonian for deformed graphene is derived. 
The scattering approach to transport is briefly described in Section \ref{sec:scattering}. 
Section \ref{sec:transport} is devoted to the exact solution of the Dirac-Kronig-Penney
model for graphene with one-dimensional modulations of strain in transport direction. 
The generalization of this model is discussed at the end of the Section \ref{sec:transport}.
The results for conductance, shot-noise, and the density of states are qualitatively 
understood on the basis of the band structure analysis, which is presented in Appendix \ref{app:band}. 
In Section \ref{sec:2D} we discuss transport in the direction perpendicular to the strain modulation. 
We summarize our results in the concluding Section \ref{sec:conclusion}.

\section{Effective Hamiltonian}\label{sec:model}

The electromechanical coupling in deformed graphene membranes
has been investigated theoretically by many authors\cite{Morozov06,Guinea08,Kim08,Isacsson08} 
following earlier publications on carbon nanotubes.\cite{Kane97, Kleiner01, Sasaki05,Huertas06} 
The deformations affect the hopping integrals in the tight-binding description of graphene in two distinct ways: 
by changing the distance between carbon atoms and by tilting the electronic $p_z$-orbitals, 
which are responsible for conduction.  However, for most cases of interest, the in-plane strains 
play the major role in determining the electronic properties of deformed graphene, 
while the tilting can be neglected.

The effects of in-plane strain are well captured by the tight-binding hamiltonian 
\begin{eqnarray}
\label{model} 
\mathcal{H} = - \sum_{\boldsymbol R}\sum_{\alpha=1}^3 \left[t+\delta t_\alpha (\boldsymbol{R})\right]
\left( a^\dagger_{\boldsymbol{R} + \boldsymbol{\delta}_\alpha} b_{\boldsymbol{R}} + \text{h.c.} \right),
\end{eqnarray}
where the summation runs over the atomic positions, $\boldsymbol{R}$, of a honeycomb lattice, 
the annihilation operators $a_{\boldsymbol{R}}$ and $b_{\boldsymbol{R}}$ refer 
to the occupation of $p_z$-orbitals at the two non-equivalent positions, $A$ and $B$, 
of the unit cell, and the three vectors $\boldsymbol{\delta}_\alpha$ shown in Fig.~\ref{fig:theta}
are directed from a $B$-atom to its three nearest neighbors. 

If the strain varies smoothly on atomic distances, the deviation of the hopping integral 
from its unperturbed value $t\approx 2.7$\,eV in a perfect crystal can be parameterized as
\begin{equation}
\label{dt}
\delta t_\alpha(\boldsymbol{R}) = (\beta t/a^2) \boldsymbol{\delta}_\alpha^\text{T} \hat{u}(\boldsymbol{R}) \boldsymbol \delta_\alpha, 
\end{equation}
where $\beta = - \partial \ln t/\partial \ln a \approx 2$ and $\hat{u}(\boldsymbol{R})$
is  the strain tensor of the graphene membrane. 

The dimensionless strain tensor describes a change in a metric that can be expressed as
\begin{equation}
d\ell^2 = d\ell_0^2+2u_{ij}ds^i ds^j, 
\end{equation}
where the summation over the spatial indices $i,j=x,y$ is assumed.
The length elements $d\ell^2_0 = \delta_{ij}ds^i ds^j$ and
$d\ell^2 = g_{ij}ds^i ds^j$ correspond to the metric $\delta_{ij}$
in the flat space and to the local metric $g_{ij}$ in the membrane, respectively. 
In subsequent formulas we let conventionally $s^x=x$ and $s^y=y$.
The graphene crystal withstands very large internal strains so that 
the values of the strain tensor elements may reach $20$\%.\cite{Lee08}

Even though the strain tensor makes no reference to the crystal structure
of graphene, the lattice symmetry is entering the hopping integral in Eq.~(\ref{dt})
due to the vectors $\boldsymbol{\delta}_\alpha$, $\alpha=1,2,3$.  The slow variation of $\hat{u}(\boldsymbol{R})$
on the scale of the lattice spacing justifies the effective mass approximation,
which is formulated in a generic reference frame shown in Fig.~\ref{fig:theta}.
In this frame we find 
\begin{equation}
\boldsymbol{\delta}_\alpha = a \left(\begin{array}{c} \sin\left(\theta+\theta_\alpha\right)\\
-\cos\left(\theta+\theta_\alpha\right)\end{array}\right), \qquad \theta_\alpha =2\pi \alpha/3,
\end{equation}
while the positions of the two non-equivalent Dirac points in the reciprocal space
are given by
\begin{equation}
\boldsymbol{K}_{1,2}=\pm\frac{4\pi}{3a\sqrt{3}} \left(\begin{array}{c}\cos\theta \\ \sin\theta \end{array}\right).
\end{equation}
Using the Fourier ansatz
\begin{equation}
\label{ansatz}
a_{\boldsymbol{R}}= \sum_{s=1}^2 e^{i \boldsymbol{K}_s \boldsymbol{R} }a_{s}(\boldsymbol{R}), \quad
b_{\boldsymbol{R}}=\sum_{s=1}^2 e^{i \boldsymbol{K}_s \boldsymbol{R} }b_{s}(\boldsymbol{R}),
\end{equation}
we obtain, to the leading order in spatial gradients, the effective model
\begin{equation}
\label{ham0}
{\cal H}=\int d^2\! \boldsymbol{R}\; 
\Psi^\dagger_{\boldsymbol{R}}\, \left[-i \hbar v \boldsymbol{\sigma}\boldsymbol{\nabla} + \tau_z \boldsymbol{\sigma}\boldsymbol{A}_\theta(\boldsymbol{R}) \right]\, \Psi_{\boldsymbol{R}},
\end{equation}
where $\hbar v= 3 t a/2$, $\tau_z$ is the Pauli matrix in the valley space and 
the operators are arranged into the four-spinor 
\begin{equation}
\Psi_{\boldsymbol{R}}=\big(\begin{array}{cccc} a_1 & e^{i\theta}b_1 & -e^{-i\theta} b_2 & a_2 \end{array}\big).
\end{equation}
The vector field  $\boldsymbol{A}_\theta=(A_{\theta x},A_{\theta y})$ is real and its  
components are related to the strain tensor in a simple way,
\begin{equation}
\label{relation}
A_{\theta x} -i A_{\theta y} = (\hbar v \beta/a) e^{3i\theta} (u_{xx}+2iu_{xy}-u_{yy}).   
\end{equation}
In the derivation of the effective Hamiltonian (\ref{ham0}) we neglected 
the velocity renormalization, which would appear as a correction 
of the order $\delta t_\alpha$ to the prefactor of the spatial gradients.\cite{deJuan07}

Unlike the usual vector potential, the vector field $\boldsymbol{A}_\theta$ 
preserves the time reversal symmetry of the Hamiltonian (\ref{ham0}). 
It also exhibits the discrete rotational invariance of the honeycomb lattice. 
Indeed, the Hamiltonian (\ref{ham0}) remains invariant under the rotation through the angle 
$\theta=2\pi/3$, while the rotation through the angle $\theta=\pi/3$ is equivalent 
to an interchange of valleys. 

The effect of a constant uniaxial strain has been studied both
theoretically\cite{Pereira2009b,Fogler08} and experimentally\cite{Kim09,Mohiuddin09} and falls beyond the scope of
our consideration. If the constant uniaxial strain (in $x$ direction) exceeds a certain
critical value, the Dirac points merge and a band gap opens.   It is predicted\cite{Pereira2009b} that, for a
crystal expanded uniformly in the zigzag direction ($\theta=0$), the
critical expansion takes on its minimal value ($\approx 23\%$), while uniaxial strain in armchair direction, $\theta=\pi/2$, never generates a gap. These predictions await experimental verification.

The strain also induces an electrostatic potential due to a change of
the on-site energies of the tight-binding model. This effect leads to the
appearance of a scalar electrostatic potential, which has to be
added to the effective Hamiltonian (\ref{ham0}).

\section{Scattering approach}
\label{sec:scattering}

In this Section we formulate the scattering approach to transport through a
deformed graphene sample with metallic leads. We take advantage of  
the effective single-particle Hamiltonian 
\begin{equation}
\label{dirac}
H=-i \boldsymbol{\sigma}\boldsymbol{\nabla} + \tau_z\boldsymbol{\sigma}\boldsymbol{A}_\theta + V, 
\end{equation}
where the fictitious vector potential $\boldsymbol{A}_\theta$ is related to 
the strain tensor, $\hat{u}(\boldsymbol{R})$, by means of the relation (\ref{relation}),
and the scalar field $V(\boldsymbol{R})$ describes strain-induced and external electrostatic potentials.
In most of the intermediate expressions we let $\hbar v=1$ for simplicity. 

The metal leads are modeled by letting $V=-U_{\textrm{lead}}$ for $x<0$ 
and $x>L$,\cite{Tworzydlo06} in Eq.~(\ref{dirac}). The vector potential 
is assumed to be zero in the leads. The width of the sample 
in $y$ direction is denoted as $W$. The scattering off the metal leads 
are fully taken into account in the subsequent analysis. 

The rectangular sample geometry makes it convenient 
to employ the Fourier transform in $y$,
\begin{equation}
\Psi_{\boldsymbol{R}}=\sum\limits_{q} e^{i q y}\Psi_{q}(x),
\end{equation}
where $q$ is the quasiparticle momentum component parallel 
to the graphene-metal interface. The momentum takes on the 
quantized values, $q=q_n$, which depend on the
boundary conditions in $y$ direction, for example, 
$q_n=2\pi n/W$ with integer $n$ for periodic boundary conditions.
In the limit $W\gg L$, which we assume below, the particular 
type of the boundary conditions is not important. 

We also restrict our consideration to small energies, $|\varepsilon| \ll U_{\textrm{lead}}$, 
where the energy $\varepsilon$ is measured with respect to the Dirac point. 
In this approximation we derive the equation on the transfer matrix in the form\cite{Titov07}
\begin{equation}
\label{Meq0}
\frac{\partial {\cal M}}{\partial x} = \left[\sigma_x (\hat{q}+\tau_z \hat{A}_{\theta y}) 
+ i\sigma_z(\hat{V}\!-\!\varepsilon) + i \tau_z \hat{A}_{\theta x}\right]{\cal M},
\end{equation}
where we introduced matrix notation in Fourier (channel) space, e.g.
\begin{equation}
\hat{V}_{nm}(x)=\frac{1}{W}\int_0^W\!\!\! dy\,e^{i(q_n-q_m)y}\,V(x,y),
\end{equation}
and $\hat{q}_{nm}=\delta_{nm}q_n$. 
The transfer matrix has the following structure in $\sigma$-space
\begin{equation}
\label{Mmatrix}
{\cal M} =\begin{pmatrix} 1/\hat{t}'\phantom{}^\dagger & \hat{r} \hat{t}^{-1} 
\\ -\hat{t}^{-1}\hat{r}' & 1/\hat{t} \end{pmatrix},
\end{equation}
where $\hat{r}$($\hat{r}'$) for $x=L$ is the matrix of reflection amplitudes 
for quasiparticles entering the sample from the left(right) lead. 
The matrices $\hat{t}$ and $\hat{t}'$ contain the corresponding transmission amplitudes.
Then, the Landauer formula for the conductance can be cast in the following form
\begin{equation}
G=\frac{2e^2}{h} {\mathrm{Tr}}\, \hat{t}\hat{t}^\dagger = 
\frac{2e^2}{h}{\mathrm{Tr}}\, \left[\hat{{\cal M}}_{11} \hat{{\cal M}}_{11}^\dagger\right]^{-1},
\end{equation}
where the trace is referred to the valley and channel space
and the symbol $\hat{{\cal M}}_{11}$ stands for the 
$11$ block of the transfer matrix in $\sigma$-space.

The Hamiltonian (\ref{dirac}) with vanishing electrostatic potential, $V=0$, 
obeys the chiral symmetry $\sigma_z H \sigma_z = - H$, which is responsible 
for a non-Abelian Aharonov-Casher gauge invariance at zero energy,\cite{Aharonov79}
\begin{equation}
\label{gauge}
\Psi'_0 = e^{i\phi +\tau_z \sigma_z \chi} \Psi_0.
\end{equation}
The spatially dependent phases $\phi$ and $\chi$ can be chosen in such a way that
the zero-energy spectral equation $H\Psi_0=0$ is reduced to the Dirac equation, 
$-i\boldsymbol{\sigma\nabla} \Psi'_0 =0$, with zero vector potential. This gauge transformation 
can be applied in the scattering approach in order to demonstrate 
that the presence of arbitrary vector potential has no effect on charge transport 
at the Dirac point as far as the contribution from edge states
can be disregarded.\cite{Schuessler09, Hannes10}

\section{1D Dirac-Kronig-Penney model}
\label{sec:transport}
\subsection{Transport}

One-dimensional modulations of strain were realized experimentally in suspended 
graphene films using the remarkably large and negative thermal expansion 
of graphene.\cite{Bao09} Motivated by these experiments we calculate
the transport properties of the one-dimensional Dirac-Kronig-Penney model, which has
been introduced earlier by several authors.\cite{Masir09,Barbier10,Arovas10} 
Below we consider a general form of the model, where the 
variation of both electrostatic as well as vector potentials is included.
Using the scattering approach formulated in the Section \ref{sec:scattering} 
we provide simple analytical solutions for transport and density of states 
in this model, which complement previous theoretical studies of Dirac Fermions 
in periodic potentials.\cite{Park08a, Park08b,DellAnna09,Snyman2009,Brey2009,Esmailpour10,Li2010,Tan10,Park2010,Novikov05}

A graphene sample with one-dimensional modulation of strain is
characterized only by the component $u_{xx}(x)$ of the strain tensor,
which depends solely on the $x$-coordinate. 
From Eq.~(\ref{relation}), the components of the pseudo-vector potential take the form
\begin{equation}
\label{Afromu}
A_{\theta x}=\frac{\beta u_{xx}}{a}  \cos 3\theta, \qquad A_{\theta y}=
-\frac{\beta u_{xx}}{a}  \sin 3\theta\,.
\end{equation}
In addition, the strain induces a spatial variation of electrostatic potential,  $V=V(x)$.
Its relation to the strain tensor is, however, complicated by screening effects.

Since the potentials depend only on $x$, the transverse momentum $q$ is conserved. 
In this case the matrices $\hat{V}$, $\hat{A}_{\theta x}$ and $\hat{A}_{\theta y}$ in Eq.~(\ref{Meq0}) 
are diagonal in channel space. Since the considered potentials also do not couple the valleys, 
the scattering problem is reduced to the solution of $2\times 2$ matrix equation,
\begin{equation}
\label{Meq}
\frac{\partial {\cal M}}{\partial x} = \left[\sigma_x (q+ A_{\theta y}) 
- i\sigma_z (\varepsilon-V) + i A_{\theta x}\right]{\cal M}.
\end{equation}
The $x$-component of the vector potential, $A_{\theta x}$, enters the equation 
in a trivial way and can be excluded by the gauge transformation 
${\cal M} \to {\cal M} \exp(i \int_0^x A_{\theta x} dx')$, 
which does not affect any observable. The first and
the most trivial consequence of this transformation 
is that the one-dimensional strain modulations in $x$ direction 
have no effect on transport in the zigzag direction ($\theta= 2 \pi n/3$ with integer $n$)
provided the effect of $V$ is ignored. 

For other angles the one-dimensional strain modulations lead to the
appearance of new minima in the gate voltage dependence of the
conductivity of the graphene sample.  The numerical solution of
Eq.~(\ref{Meq}) suggests that a periodic vector potential induces much
more pronounced minima than a periodic scalar potential of equivalent
amplitude. The weaker effect of $V$ can be associated with Klein
tunnelling, which leads to the suppression of pseudo-gaps in the
spectrum of the superstructure. 

In order to describe the effect of periodic potentials analytically 
we introduce the one-dimensional Dirac-Kronig-Penney model, 
which is characterized by the vector potential $\boldsymbol{A}$ given by Eq.~(\ref{Afromu}) with
\begin{equation}
\label{Kronig_u}
u_{xx}(x) = -u_0/2+u_0\sum_j \Theta(x-2\ell j)\, \Theta((2j\!+\!1)\ell\!-\!x),
\end{equation}
where the scale $\ell$ stands for half the period of the superlattice, 
$\Theta(x)$ is the Heaviside step function, and the dimensionless parameter 
$u_0$ specifies the amplitude of the strain modulation. 
The vector potential introduced by Eqs.~(\ref{Afromu},\ref{Kronig_u}) corresponds to a strain field which is smooth on atomic scale but changes abruptly on distances smaller than the Dirac quasiparticle wave length,  $\hbar v/\varepsilon$. The periodic electrostatic potential is introduced in a similar manner,
\begin{equation}
\label{Kronig_v}
V(x) = -V_0/2+V_0\sum_j \Theta(x-2\ell j)\, \Theta((2j\!+\!1)\ell\!-\!x),
\end{equation}
where $V_0$ is the amplitude of the modulation. 

 
\begin{figure}[tb]
\centerline{\includegraphics[width=0.9\columnwidth]{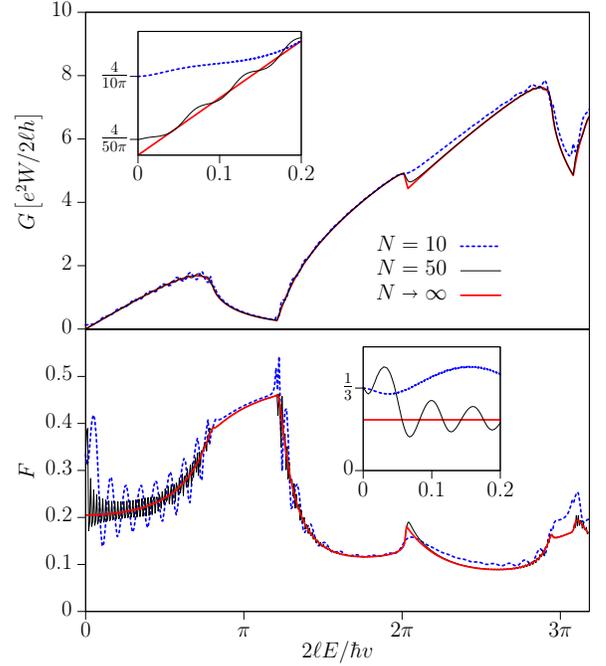}}
\caption{(Color online) Conductance and Fano factor calculated from Eq.~(\ref{Gexact}) 
for finite systems comprising $N=10$ and $N=50$ supercells are 
compared with the averaged expressions to be derived in Eqs.~(\ref{Gav},\ref{Fav}),
which correspond to the limit $N\to \infty$ and ignore contributions 
from evanescent modes. The plots are for graphene 
with strain modulations, $r_{\theta x} = 1$. 
In a small vicinity of the Dirac point, $\varepsilon \ll \hbar v/L$, 
(shown at the insets) the transport is insensitive 
to strain due to the extended gauge invariance (\ref{gauge}) 
and is dominated by the evanescent modes.
}
\label{fig:Ncompare}
\end{figure}

For the sake of simplicity we consider a sample of length $L=2\ell N$ 
with $N$ being an integer. In this case the solution to Eq.~(\ref{Meq}) 
satisfies ${\cal M}(L)={\cal M}_0^N$, where ${\cal M}_0$ is 
the transfer matrix corresponding to the size of the supercell, $2\ell$. 
Disregarding the $x$-component of the vector potential, $A_{\theta x}$,  
which has been argued above to have no effect on transport, we write
\begin{equation}
\label{M0}
{\cal M}_0={\cal M}_+{\cal M}_-,\qquad {\cal M}_{\pm}=e^{(q_\pm\sigma_x-i\varepsilon_{\pm}\sigma_z)\ell},
\end{equation}
where we take advantage of the definitions
\begin{equation}
q_\pm=q \pm (\beta u_0/2a)\sin 3\theta, \qquad \varepsilon_\pm=\varepsilon\pm V_0/2.
\end{equation}
The eigenvalues of the transfer matrix, ${\cal M}_0$, 
are conveniently parameterized as
\begin{equation}
\label{k0}
e^{\pm 2 i k_0\ell} =\lambda\pm i \sqrt{1-\lambda^2},
\end{equation}
where we introduced the real function 
\begin{equation}
\lambda = \cos k_+\ell\,\cos k_-\ell +\frac{q_+q_-\!-\!\varepsilon_+\varepsilon_-}{k_+k_-} \sin k_+\ell\,\sin k_-\ell, 
\label{lambda}
\end{equation}
with $k_{\pm}=(\varepsilon_{\pm}^2-q_{\pm}^2)^{1/2}$ the $x$-component of the momenta. 
The wave number $k_0$ becomes imaginary for some values 
of $q$ and $\varepsilon$ indicating the appearance of pseudo-gaps 
in the superstructure. In the considered Dirac-Kronig-Penney model the value of $k_0$
is nothing but the $x$-component of the quasi-momentum in the reciprocal space
associated with the superstructure.

In order to calculate the element ${\cal M}_{11}$ of the full transfer-matrix 
${\cal M}(L)$ we use the Chebyshev identity to calculate the $N$-th power 
of an unimodular matrix\cite{Born99}
\begin{equation}
\label{Chebyshev}
{\cal M}_0^N=
\frac{{\cal M}_0\sin k_0L -  \sin k_0(L\!-\!2\ell)}{\sin 2k_0\ell}.
\end{equation}
The matrix element $\left({\cal M}_0\right)_{11}=\lambda-i\eta$ is readily determined from 
Eqs.~(\ref{M0},\ref{Chebyshev}) with
\begin{equation}
\label{eta}
\eta=\frac{\varepsilon_-}{k_-}\sin k_-\ell\, \cos k_+\ell+\frac{\varepsilon_+}{k_+}\sin k_+\ell \,\cos k_-\ell.
\end{equation}
Using the fact that both $\lambda$ and $\eta$ are real functions of the energy $\varepsilon$,
and the conserved momentum component $q$, we obtain from Eqs.~(\ref{Mmatrix},\ref{Chebyshev}) 
the exact transmission probabilities for each value of $q$,
\begin{equation}
\label{exact}
T_q=\left[\cos^2{k_0L}+\frac{\eta^2}{1-\lambda^2} \sin^2{k_0L}\right]^{-1}.
\end{equation}
This expression is reduced to the well-known result for a purely ballistic system,\cite{Tworzydlo06} i.e. 
for $u_0=V_0=0$, by the substitution $\eta=(\varepsilon/k)\sin 2k\ell$ and $\lambda =\cos 2k\ell$. 
The transmission probabilities (\ref{exact}) determine the energy-dependent transport quantities
\begin{equation}
\label{Gexact}
G=\frac{4 e^2}{h}\sum\limits_q T_q, \qquad F=\frac{\sum_q T_q(1-T_q)}{\sum_q T_q},
\end{equation}
where $G$ is the Landauer conductance and $F$ is the Fano factor for the shot noise.
In the limit $W\gg L$ the summation over the scattering channels
can be replaced by the integration,  $\sum_q\to (W/2\pi)\int dq$.

\begin{figure}[tb]
\centerline{\includegraphics[width=0.9\columnwidth]{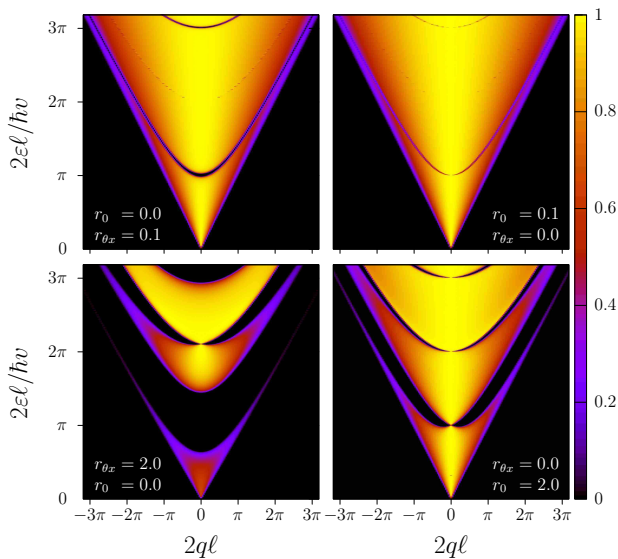}}
\caption{The density plot for the averaged transmission spectra, $\overline{T}_q(\varepsilon)$, 
calculated from Eq.~(\ref{Tav}) for the Dirac-Kronig-Penney model with strain modulation
$r_{\theta x}=0.1$ (top left), $r_{\theta x} = 2.0$ (bottom left) 
and electrostatic field modulation $r_0=0.1$ (top right),  $r_0=2.0$ (bottom right).
}
\label{fig:Tdensity}
\end{figure}

More generally one can define the cumulant generating function for the transport,
\begin{equation}
\label{full}
{\cal F}(\chi)=\sum\limits_q \ln(1-T_q+e^{\chi}T_q).
\end{equation}
The dimensionless cumulants $c_n= \lim_{\chi\to0}\partial^n {\cal F}/{\partial \chi^n}$
determine the so-called full counting statistics of the charge transport.  The conductance 
and the Fano factor are given by $G=(4e^2/h)c_1$ and $F=c_2/c_1$, respectively.

The strength of strain-induced and electrostatic potentials in the Dirac-Kronig-Penney model 
(\ref{Kronig_u},\ref{Kronig_v}) is characterized by the dimensionless parameters
\begin{equation}
\label{dimensionless}
r_{\theta x}=(2 u_0 \ell /a ) \sin 3\theta, \qquad r_0 = \ell V_0/\hbar v,
\end{equation}
respectively. In Fig.~\ref{fig:Ncompare} the conductance and the Fano factor in graphene 
with periodic modulations of strain calculated from Eq.~\eqref{Gexact}  
are plotted for systems with finite lengths $L=20 \ell$ and $L=100 \ell$ for $r_{\theta x}=1$.

The energy dependence of conductance and Fano-factor (\ref{Gexact}) reveal 
fast Fabry-P\'erot oscillations on the scale $\hbar v/L$. From a physics point of view, 
these oscillations originate from multiple reflections of propagating modes (real $k_0$)
at the metal-graphene interfaces. 

The channels with imaginary values of $k_0$ (evanescent modes or metal-induced states) 
also contribute to transport. 
Even though the individual contribution of each evanescent mode is exponentially small, 
$T_q\sim \exp( -2L {\mathrm{Im}}\, k_0)$, their combined effect becomes essential for energies 
in a vicinity of band-edges. The role of evanescent modes is especially important at the Dirac point
due to the absence of propagating modes. Indeed, 
the conductance and the Fano-factor determined by (\ref{Gexact}) take on the values
$G=4e^2 W/\pi h L$ and $F=1/3$ for $\varepsilon\ll \hbar v/L$ irrespective of the vector potential. 
This universality is due to the extended gauge invariance (\ref{gauge}).

\begin{figure}[tb]
\centerline{\includegraphics[width=0.8\columnwidth]{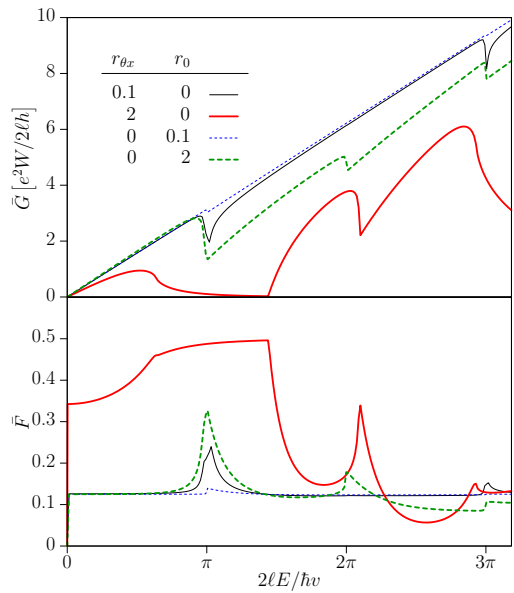}}
\caption{(Color online) The averaged conductance and Fano factor calculated from Eqs.~(\ref{Gav},\ref{Fav})
for the one-dimensional Dirac-Kronig-Penney model with different field strengths (\ref{dimensionless}). 
The effect of the scalar potential is suppressed due to Klein tunnelling through potential barriers.}
\label{fig:GFstrength}
\end{figure}

For $N=L/2\ell \gg 1$, the amplitude of the Fabry-P\'erot oscillations in $G$ and $F$ decreases 
(provided $W\gg L$) and the relative contribution of evanescent modes becomes less important.
Taking the limit $N\to \infty$ is equivalent to ignoring the imaginary values of $k_0$ and averaging over the rapid phase, $k_0 L$. 
This approximation has been used e.g. in Ref.~\onlinecite{Hannes10} to obtain the full counting statistics of few-layer graphene. 

The averaged generating function (\ref{full}) takes the form
\begin{equation}
\label{fullav}
\overline{\cal F}(\chi)=2\sum\limits_q \ln \left(e^{\chi/2}+\sqrt{\overline{T}_q^{\,-2}+e^{\chi}-1}\,\right),
\end{equation}
where we introduce the mean transmission probability
\begin{equation}
\label{Tav}
\overline{T}_q=|\eta|^{-1}{\mathrm{Re}}\, \sqrt{1-\lambda^2}.
\end{equation}
The averaged transport quantities do not depend on the system size $L$ and reveal 
no Fabry-P{\'e}rot oscillations. For a ballistic sample, $u_0=V_0=0$, one finds 
$\overline{T}^\textrm{ball}_q=k/|\varepsilon|$, where $k=\sqrt{\varepsilon^2-q^2}$ is the $x$-component of the momentum.

\begin{figure}[tb]
\centerline{\includegraphics[width=0.9\columnwidth]{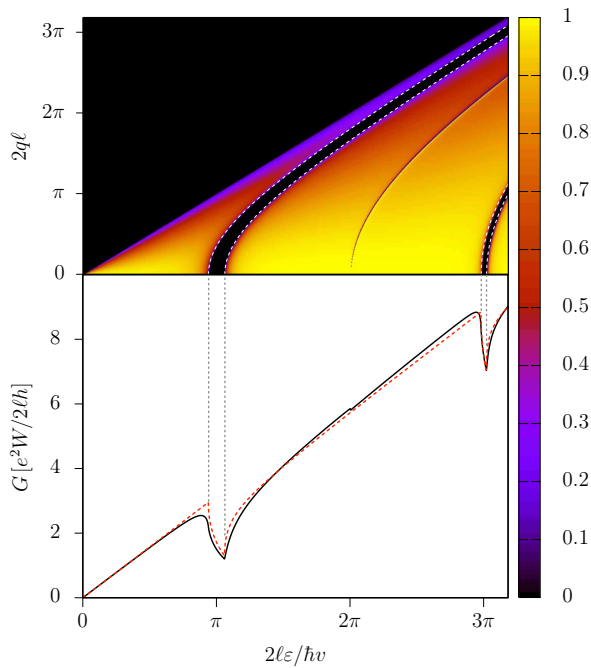}}
\caption{(Color online) The averaged conductance of Eq.~(\ref{Gav}) (solid) is compared with an estimate (\ref{modecounting}) 
based on the mode-counting argument (dashed) for the Dirac-Kronig-Penney model with $r_{\theta x}=0.3, r_0=0$.
The upper panel shows the corresponding transmission spectrum.}
\label{fig:ModeCounting}
\end{figure}

From the generating function (\ref{fullav}) we readily find the averaged conductance and noise.
The conductance is given by the Landauer formula,
\begin{equation}
\label{Gav}
\overline{G}=\frac{4e^2}{h}\sum\limits_q \overline{T}_q,
\end{equation}
while the Fano-factor is related to the averaged transmission probabilities 
in a less evident way
\begin{equation}
\label{Fav}
\overline{F}=\frac{\sum_q \overline{T}_q\left(1-\overline{T}_q^2\right)}{2\sum_q \overline{T}_q}.
\end{equation} 

The exact and averaged transmission probabilities (\ref{exact},\ref{Tav}) 
together with the corresponding expressions for the full counting statistics
(\ref{full},\ref{Fav}) provide the complete analytical description of the transport properties
in the 1D-Dirac-Kronig-Penney model with scalar and vector potentials of arbitrary strength. 

The dependence of the transmission coefficient on the transversal
momentum component, $q$, is called the transmission spectrum. 
In Fig.~\ref{fig:Tdensity} we plot the transmission spectra 
obtained from Eq.~(\ref{Tav}) for different strengths of the scalar and vector potentials. 

\begin{figure}[tb]
\centerline{\includegraphics[width=0.9\columnwidth]{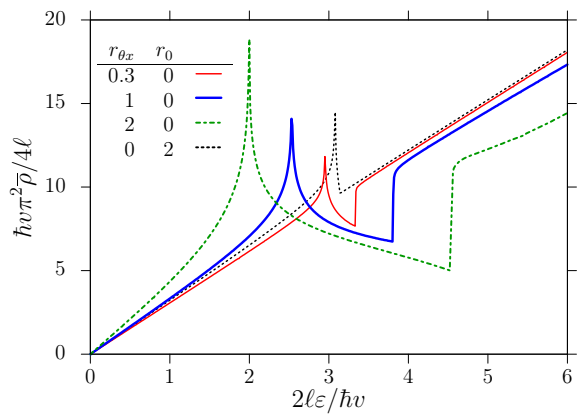}}
\caption{(Color online) The averaged density of states obtained from Eq.~(\ref{rhoav}) for the Dirac-Kronig-Penney
model with different strengths of the periodic potentials. }
\label{fig:dos}
\end{figure}

The transmission spectra shown in Fig.~\ref{fig:Tdensity} are qualitatively different 
for weak, $r_{\theta x}\ll 1$, and strong, $r_{\theta x} \gg 1$, strain modulations. 
It has to be stressed that the latter regime is well within the experimentally accessible range of parameters. 
Indeed, for an achievable strain modulation of $0.2$\% ($u_0=0.002$) and the period $\ell=70$\,nm 
one finds $r_{\theta x} \approx 2$.

Neither the modulated strain nor the electrostatic field does open up a full band gap for any values
of the parameters $r_{\theta x}$ or $r_0$. Nevertheless, very large pseudo-gaps are generated 
by strain modulations with $r_{\theta x}\gg 1$. One can see from the lower left panel 
in Fig.~\ref{fig:Tdensity} that the charge transport in this regime is suppressed 
for almost all directions of the momenta (different $q$ values)  in a wide energy range. 
In contrast, the pseudo-gaps do not open at normal incidence ($q=0$) in the Dirac-Kronig-Penney
model with periodic electrostatic potential due to the Klein-tunnelling of quasiparticles 
through electrostatic barriers.\cite{Katsnelson06}

The effects of strain and electrostatic field modulations on the conductance and 
noise are compared in Fig.~\ref{fig:GFstrength}. The most prominent feature of 
these plots are the dips in conductivity centered around the energies 
$\varepsilon_n= n \pi \hbar v /2\ell$ with $n=1,2,3,\dots$. 
The physical mechanism, which is responsible for the appearance of the pseudo-gaps, 
is equivalent to that of the parametric resonance.\cite{Arnold89}

For $r_{\theta x}\gg 1$, the conductance is suppressed in wider energy intervals around $\varepsilon=\varepsilon_n$. 
For weak strain or potential amplitude, $r_{\theta x}\ll 1$ or $r_0\ll 1$, significant pseudo-gaps
are only located around the energy values $\varepsilon_n$ with odd values of $n$, since 
only odd Fourier components of the potentials (\ref{Kronig_u},\ref{Kronig_v}) exist.
For a weak harmonic potential, a significant pseudo-gap would only arise around $\varepsilon=\varepsilon_1$.

The areas of vanishing transmission probability in Fig.~\ref{fig:Tdensity} (top left) can be understood on the basis of perturbation theory as presented in Appendix~\ref{app:band} for small amplitudes of periodic potentials. 
The pseudo-gap emerging due to a weak periodic strain is described 
by the functions $\varepsilon_{g-}(q)$ and $\varepsilon_{g+}(q)$ given in the first row of Tab.~\ref{tab:gaps}. 
These functions are shown with dashed lines in the upper panel of Fig.~\ref{fig:ModeCounting}. 

Inside the pseudo-gap, i.e. for $\varepsilon_{g-}(q)<\varepsilon<\varepsilon_{g+}(q)$, the transmission coefficient
is exponentially small, while it is close to the ballistic value, $\overline{T}^\textrm{ball}_q=\sqrt{\varepsilon^2-q^2}/|\varepsilon|$, 
outside the pseudo-gap (one can disregard the change in the resistance of the metal-graphene interface 
due to the weak periodic potentials). Note further, that the pseudo-gap is located near $q=0$ for energies near the dip such that $\overline{T}^\textrm{ball}_q\approx 1$ in this region.  We can, therefore, estimate the conductance in the vicinity of the lowest dip ($\varepsilon=\varepsilon_1$) by subtracting the contribution 
of ballistically propagating modes inside the pseudo-gap from the ballistic result as
\begin{equation}
\label{modecounting}
G_{mc} =  G_0 - \frac{4e^2}{h}\sum\limits_q 
\Theta(\varepsilon\!-\!\varepsilon_{g-}(q))\,\Theta(\varepsilon_{g+}(q)\!-\!\varepsilon),
\end{equation}
where $\Theta(x)$ is the Heaviside step function, $G_0=e^2 W\varepsilon/h$, and the limit $W |\varepsilon |\gg 1$ is assumed. The generalisation of Eq.~\eqref{modecounting} around the higher resonant energies, $\varepsilon=\varepsilon_n, n=3, 5, \dots$, is straightforward.

It is shown in Fig.~\ref{fig:ModeCounting} for $r_{\theta x}=0.3$ that the result of Eq.~(\ref{modecounting}) 
agrees with the averaged conductance calculated from Eq.~(\ref{Gav}). The band-structure analysis 
(\ref{modecounting},\ref{secular}) predicts characteristic dips in the conductance at 
$\varepsilon=n\pi \hbar v/2\ell, n=1,3,5\dots$ of the depth $\delta G = (8e^2 W/h \pi)\sqrt{|A_n|n\pi/2\ell}$, 
where $A_n$ stands for $n$-th Fourier-component of the vector potential. 
For the Dirac-Kronig-Penney model one finds $A_n\propto 1/n$ hence the value of the conductance 
dip does not depend on $n$. We note that the validity of Eq.~(\ref{modecounting}) is restricted to 
weak potentials. For $r_{\theta x}, r_0 \gtrsim 1$, the pseudo-gaps overlap and 
the perturbation theory of Appendix~\ref{app:band} is no longer applicable.

\subsection{Density of states}

Let us now extend the solution of the Dirac-Kronig-Penney model to 
the density of states. For purely ballistic system with metal leads, both the local and the integrated 
density has been found in Ref.~\onlinecite{Titov10} using a Green's function approach. 
Below we take advantage of an alternative route and relate the partial density of states 
in the channel $q$ to the corresponding transmission amplitude, $t_q$, 
by the well-known formula\cite{Buettiker93} 
\begin{equation}
\nu_q=-\frac{1}{\pi} {\mathrm{Im}}\left[ \frac{\partial  \ln t_q}{\partial  \varepsilon}\right].
\end{equation}
Then, the integrated density of states per unit volume is given by
\begin{equation}
\label{IDOS}
\rho(\varepsilon)=\frac{4}{L W}\sum\limits_q \nu_q(\varepsilon), 
\end{equation}

where the factor $4$ takes into account the spin and valley degeneracy.

The transmission amplitude is readily obtained from Eqs.~(\ref{Mmatrix},\ref{Chebyshev}) as
\begin{equation}
t_q=\frac{1}{\cos{k_0L}+i \zeta_q \sin{k_0L}},\qquad \zeta_q=\frac{\eta}{\sin 2k_0\ell},
\end{equation}
where the wave-number $k_0$ and the real quantity $\eta$ are determined from the 
expressions (\ref{k0}) and (\ref{eta}), respectively. Therefore, the partial density of states 
can be calculated exactly as
\begin{equation}
\label{nuexact}
\nu_q=
\frac{L \zeta_q (dk_0/d\varepsilon) + \sin k_0L\,\cos k_0L\, (d \zeta_q/d\varepsilon)}
{\pi\left(\cos^2k_0L+\zeta_q^2 \sin^2 k_0 L\right)}.
\end{equation}
Unlike the conductance or the shot noise (\ref{Gexact}), the density of states depends 
on the value of $U_{\textrm{lead}}$, because the spectrum in a vicinity 
of the band-edges is dominated by the metal-induced states (evanescent modes). 

The metal proximity effect\cite{Titov10}  can be seen already for purely 
ballistic system near the Dirac point. Indeed, for $u_0=V_0=0$,
one finds
\begin{equation}
\label{ball}
\nu^\textrm{ball}_q(\varepsilon)=\frac{L\varepsilon^2-(q^2/k)\cos kL \,\sin kL }{\pi(\varepsilon^2-q^2\cos^2 kL)},
\end{equation}
where $k=\sqrt{\varepsilon^2-q^2}$.   At zero energy the ballistic result (\ref{ball})
is reduced to $\nu^\textrm{ball}_q(0)=\tanh{(qL)}/\pi q$. 
Therefore, the density of states at $\varepsilon=0$ in Eq.~(\ref{IDOS}) acquires 
a logarithmic divergency. This divergency is regularized by the largest available
transversal momentum $q_{\textrm{max}}$, which is simply equal to 
the Fermi-momentum in the metal lead, $q_{\textrm{max}} = U_\textrm{lead}/\hbar v$.
Thus, the density of states in a close vicinity of the Dirac point, $\varepsilon\ll \hbar v/L$, 
is given by 
\begin{equation}
\rho^\textrm{ball}(0) = \frac{4}{\pi^2\hbar v L}\ln\frac{U_\textrm{lead}L}{\hbar v},
\end{equation}
in agreement with Ref.~\onlinecite{Titov10}. Similar logarithmic dependence
of the density of states on $U_\textrm{lead}$ takes place in the Dirac-Kronig-Penney model
for energies inside the pseudo-gaps. 

In full analogy with the averaged generating function (\ref{fullav}) 
we can introduce the averaged density of states, $\overline{\rho}$,
which corresponds to the limit $L\to \infty$.  In this limit we disregard the
contribution of the metal-induced states by projecting 
on the real values of the momentum $k_0$ and average the result of Eq.~(\ref{ball})
over the rapid phase $k_0L$. This procedure leads to the simple result
\begin{equation}
\label{nuav}
\overline{\nu}_q =\frac{L}{\pi} \left|\frac{d {\mathrm{Re}}\, k_0}{d\varepsilon}\right|,
\end{equation}
hence, in the limit $W\gg L$, the mean density is given by
\begin{equation}
\label{rhoav}
\overline{\rho}=\frac{2}{\pi^2}\int_{-\infty}^{\infty} \!\!\! dq\,\left|\frac{d {\mathrm{Re}}\, k_0}{d\varepsilon}\right|.
\end{equation}
In the ballistic limit, $u_0=V_0=0$, Eq.~(\ref{rhoav}) yields the density of states
of the clean graphene $\overline{\rho}^\textrm{ball}=2|\varepsilon|/\pi\hbar^2v^2$.
In Fig.~\ref{fig:dos} we plot the averaged density of states calculated 
from Eq.~(\ref{rhoav}) for different strengths of the periodic potentials.

\begin{figure}[tb]
\centerline{\includegraphics[width=\columnwidth]{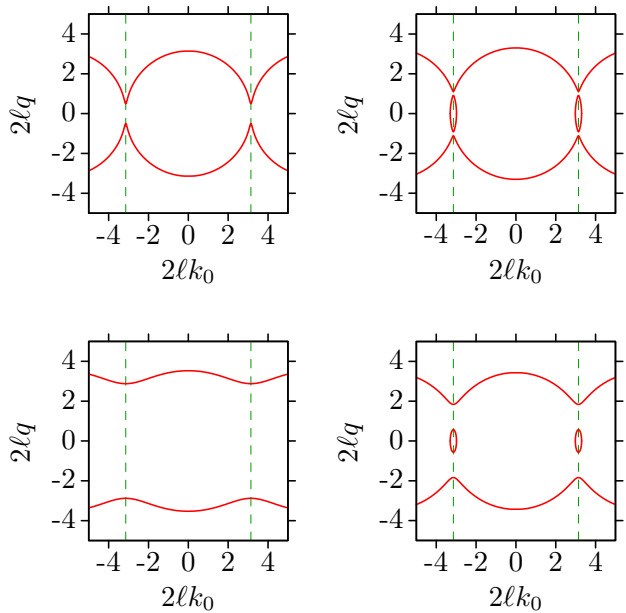}}
\caption{(Color online) The contour plot of the dispersion relation for the Dirac-Kronig-Penney
model (\ref{Kronig_u},\ref{Kronig_v}) with strain modulation
$r_{\theta x}=0.3$ (top left), $r_{\theta x} = 2.0$ (bottom left) 
and electrostatic field modulation $r_0=0.3$ (top right),  $r_0=2.0$ (bottom right) at energy $\varepsilon=3.3 \hbar v/2\ell$.}
\label{fig:FS}
\end{figure}

\subsection{Beyond the Dirac-Kronig-Penney model}

The results of the previous subsections are readily generalized to a model with arbitrary 
periodic variation of strain and electrostatic potential in $x$ direction. The analysis of the model
is reduced to the calculation of the transfer matrix, ${\cal M}_0$, which corresponds 
to the wave propagation over the distance $2\ell$, the period of the potential. 
In this generalized model, both the exact and the averaged full counting statistics 
as well as the density of states
are still given by the expressions (\ref{full},\ref{fullav},\ref{nuexact},\ref{nuav}) with 
\begin{equation}
\label{num}
\lambda=\frac{1}{2}{\mathrm{Tr}}\,{\cal M}_0,\qquad 
\eta=\frac{i}{2}{\mathrm{Tr}}\,\sigma_z{\cal M}_0,
\end{equation}

and the $x$-component, $k_0$, of the quasi-momentum is related to $\lambda$ by Eq.~(\ref{k0}).
Thus, for a periodic  potential of a general type, the full solution of the problem is
reduced to the straightforward numerical evaluation of the functions $\lambda$ and $\eta$.
Note, that the exact analytical expressions (\ref{lambda},\ref{eta}) are restricted to 
the Dirac-Kronig-Penney model and do not apply generally.  

We have used Eqs.~(\ref{num}) to calculate the energy-dependent conductance for different amplitudes
of the periodic strain in the harmonic potential, $u_{xx}=u_0 \sin(\pi x/\ell)$. In this  case, 
the higher pseudo-gaps are found to be suppressed for weak potentials $r_{\theta x},r_0 \ll 1$ 
as compared to the Dirac-Kronig-Penney model. 
The lowest conductance minima around $\varepsilon=\varepsilon_0$ is essentially the same in both models.  
The models become even more similar with increasing potential strength. 
For $r_{\theta x},r_0 \gtrsim 1$ the transmission spectra of the single-harmonic model,  
becomes almost equivalent to those of the Dirac-Kronig-Penney model for all energies.

\begin{figure}[tb]
\includegraphics[width=\linewidth]{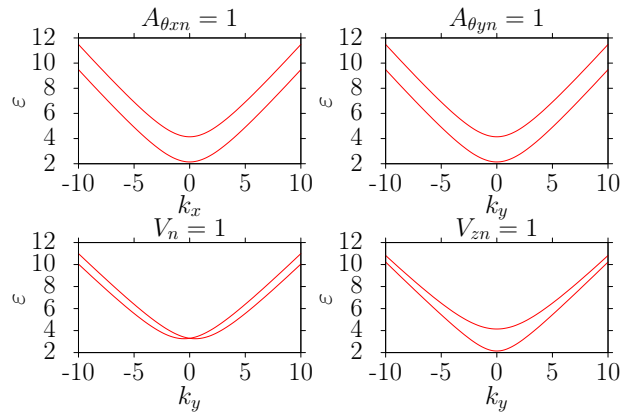}
\caption{(Color online) Typical band gap boundaries. In all plots $n, b$ was set to 1.}
\label{fig:boundaries} 
\end{figure}

\subsection{Transport in $y$ direction}
\label{sec:2D}

We have demonstrated that periodic and $x$-dependent scalar and vector potentials modify
transport in $x$-direction due to the appearance of pseudo-gaps. Let us now argue that such potentials
hardly affect transport in $y$ direction, i.e. in the direction parallel to the equi-potential lines.

In Fig.~\ref{fig:FS} we plot the Fermi surface slightly above the energy value $\varepsilon=\varepsilon_0$ 
using the exact dispersion relation, $\varepsilon=\varepsilon(k_0,q)$, obtained from Eqs.~(\ref{k0},\ref{lambda})
for the Dirac-Kronig-Penney model.  We have seen that the transport in $x$ direction is dominated by the 
modes with the momenta $q \in (-\varepsilon,\varepsilon)$. For some values of $q$ in this interval one cannot 
find any propagating states,  which correspond to the real values of $k_0$. This can be seen as 
the formation of pseudo-gaps in the transmission spectra shown in Fig.~\ref{fig:Tdensity}, which
is especially strong for $r_{\theta x},r_0\gtrsim 1$. In contrast, for each real value of $k_0$ 
one always finds real values of $q$. Therefore, the transport in $y$-direction is not affected 
by the formation of the pseudo-gaps, and depends only on the details of scattering  
at the metal-graphene interfaces. The latter effect is, however, weak and falls beyond 
the scope of our consideration.  We therefore conclude that one-dimensional superlattices have a major impact only on the transport properties along the direction of their periodicity.

\section{Conclusion}
\label{sec:conclusion}

The Dirac-Kronig-Penney model has been applied in order to study phase-coherent charge
transport in periodically strained graphene samples. Using the exact relations for the band-structure 
in the considered  superlattice, we calculate the density of states and the full counting statistics 
for charge transport. The exact quantities are simplified in the thermodynamic limit 
by neglecting the contribution of evanescent modes and averaging over the Fabry-P\' erot oscillations. 
The conductance is found to be suppressed at energies corresponding to the positions of the 
pseudo-gaps in the modified spectra of the superlattice. The periodic strain and electrostatic potentials 
are argued to have the largest effect on transport in the direction perpendicular to the equipotential lines.  
The vector potentials are shown to play a greater role in confining the Dirac quasi-particles 
due to the suppression of Klein tunnelling.

\acknowledgments

We thank W.-R. Hannes for discussions. S.G. acknowledges support by
the EPSRC Scottish Doctoral Training Centre in Condensed Matter Physics.  
W. B. acknowledges support by the DFG through SFB 767 \textit{Controlled Nanosystems} , 
by the Research Initiative \textit{UltraQuantum}, and in part 
by the Project of Knowledge Innovation Program (PKIP) 
of Chinese Academy of Sciences, Grant No. KJCX2.YW.W10.

\appendix

\section{Band structure for Dirac particles in a superlattice}
\label{app:band}
In this appendix we perturbatively calculate the band structure of a graphene sample described by the single-valley Dirac equation
\begin{equation}
\left[
-i\boldsymbol\sigma\cdot \boldsymbol\nabla + \mathcal V(\boldsymbol r)
\right] \Psi(\boldsymbol r) = \varepsilon \Psi(\boldsymbol r),
\end{equation}
where 
\begin{equation}
\mathcal V(\boldsymbol r) = \left(%
\begin{array}{cc}
V(\boldsymbol r) + V_z(\boldsymbol r)& A_{\theta x}(\boldsymbol r) - i A_{\theta y}(\boldsymbol r)\\
A_{\theta x}(\boldsymbol r) + i A_{\theta y}(\boldsymbol r) & V(\boldsymbol r)- V_z(\boldsymbol r)
\end{array}
\right)
\end{equation}
is a general 1d-periodic potential that is smooth on the scale of the graphene lattice constant $a$. To do so, we generalize the standard methods of elementary band structure theory to the Dirac case with its non-trivial matrix structure. We find that for purely scalar potentials, this additional structure prevents the creation of a gap at $k_y=0$ thus making $\boldsymbol A_{\theta}$ and $V_z$ more relevant for the transport properties; this fact can be related to the chirality of the quasi-particles in graphene and the phenomenon of Klein tunnelling. 

Let $\boldsymbol b = (b_x,b_y)^\text{T}$ span the 1d-super-lattice (so that $\mathcal V(\boldsymbol r + n\boldsymbol b) = \mathcal V(\boldsymbol r)$) and $\boldsymbol g$ be the corresponding reciprocal vector fixed by the conditions $\boldsymbol g\cdot \boldsymbol b = 2\pi$ and $\boldsymbol g\parallel \boldsymbol b$. Making the Bloch ansatz
\begin{equation}
\Psi(\boldsymbol r) =\sum_n\ \psi(\boldsymbol k-n\boldsymbol g)\ e^{i(\boldsymbol k - n\boldsymbol g)\boldsymbol r}
\label{eq:bloch_ansatz}
\end{equation}
for the wave function and writing the periodic potential as a Fourier series,
$\mathcal V(\boldsymbol r) = \sum_n \mathcal V_n\ e^{in\boldsymbol g \boldsymbol r}$,
we can write down the central equation
\begin{equation}
\begin{pmatrix}-\varepsilon &\mathrm k^\ast - n\mathrm g^\ast\\\mathrm k-n\mathrm g&-\varepsilon \end{pmatrix} \psi(\boldsymbol k-n\boldsymbol g)+
\sum_{m}\mathcal V_{m-n} \psi(\boldsymbol k-m\boldsymbol g) = 0,
\label{eq:central_equation}
\end{equation}
where $\mathrm k = k_x+i k_y$ and $\mathrm g = g_x + i g_y$. The central equation \eqref{eq:central_equation} 
represents an infinite set of linear equations for the coefficients $\psi$, which can in
general only be solved numerically. The solution, however, simplifies for sufficiently weak periodic potentials. 

For $\boldsymbol k$ close to a degeneracy, $|\boldsymbol k|=|\boldsymbol k - n\boldsymbol g|\approx \varepsilon$, we can neglect 
all but the two coefficients $\psi(\boldsymbol k)$ and $\psi(\boldsymbol k-n\boldsymbol g)$ in Eq.~ \eqref{eq:central_equation}. 
The resulting linear system has the simple form
\begin{equation}
\label{secular}
\left|\begin{array}{cccc}	-\varepsilon & \mathrm k^* & V+V_{z}&A_{\theta x}-iA_{\theta y}\\
\mathrm k & -\varepsilon &A_{\theta x}+iA_{\theta y} & V-V_{z}\\
V^*+V_{z}^* &A_{\theta x}^*-iA_{\theta y}^* & -\varepsilon & \mathrm k_n^*\\
 A_{\theta x}^*+iA_{\theta y}^* &V^*-V_{z}^* & \mathrm k_n & -\varepsilon
\end{array}\right| = 0.
\end{equation}
Here, we introduced the abbreviation $\mathrm k_n=\mathrm k-n\mathrm g$ and omitted the Fourier index $n$ of the potentials for brevity. The solution of Eq.~\eqref{secular} is particularly compact when only one of the coefficients $A_{\theta x n}, A_{\theta y n}, V_n$ or $V_{zn}$ is different from zero. In this case we obtain 
\begin{eqnarray}
\nonumber
\varepsilon^2 &=&\frac{|\mathrm k|^2+|\mathrm k_n|^2}2  + |U_{n}|^2 \\
&\pm&\frac12\sqrt{(|\mathrm k|^2-|\mathrm k_n|^2)^2   + 4|U_{n}|^2|2p-n\mathrm g|^2},
\label{eq:dispersions}
\end{eqnarray}
where the pair of quantities $(U_n,p)$ take on one of the following values 
$(A_{xn},k_x)$,  $(A_{yn}, -ik_y)$, $(V_n, k)$, or $(V_{zn},0)$.
We compare the energy values of the forbidden zone's upper/lower boundary, $\varepsilon_{g\pm}$, as well as the width of the gap in forward direction for the different types of periodic potentials in Tab.~\ref{tab:gaps}. 
To calculate $\varepsilon_{g\pm}$ we have chosen $\boldsymbol k$ in Eq.~\eqref{eq:dispersions} such, that $\boldsymbol k\cdot \boldsymbol b/b = n|\mathrm g|/2$. Typical band gap boundaries are shown in Fig.~\ref{fig:boundaries}.
\begin{table}[t]
\begin{tabular}{|l|c|c|}
 \hline&$\varepsilon_{g\pm}$\phantom{\huge M} & $\Delta(k_y=0)$\\\hline
 $A_{\theta x n}\neq 0$\phantom{\huge M}&$\sqrt{k_x^2+\left|\frac{n\mathrm g}2\right|^2}\pm|A_{\theta x n}|$&$0$\\
$A_{\theta y n}\neq 0$&$\sqrt{k_y^2+\left|\frac{n\mathrm g}2\right|^2}\pm|A_{\theta y n}|$&$2|A_{\theta y n}|$\\
$V_n\neq 0$&$\sqrt{k_y^2+\left|\frac{n\mathrm g}2\right|^2+|V_{n}|^2\pm|k_y||V_{n}|}$&$0$\\
$V_{zn}\neq 0$&$\sqrt{k_y^2+\left|\frac{n\mathrm g}2\right|^2+|V_{zn}|^2\pm2|\frac{n\mathrm g}2||V_{zn}|}$&$2|V_{zn}|$\\
\hline
\end{tabular}
\caption{Energy value of the  forbidden zone's upper/lower boundary, as well as the width of the gap in forward direction for the different types of periodic potentials. In the first row,  $\boldsymbol b$ was chosen parallel to $\boldsymbol e_y$, for the remaining rows $\boldsymbol b\parallel \boldsymbol e_x$.}
\label{tab:gaps}
\end{table}

\end{document}